\documentclass[aps, prx, 10pt,
               superscriptaddress, nofootinbib, 
               amsmath, 
               twocolumn, preprintnumbers, 
               showpacs, 
               raggedbottom, 
               floatfix,
               longbibliography, 
               ]{revtex4-1}

\newcommand{\exclude}[1]{}

\usepackage[aps, midcaps, nomacros]{my_paper}

\usepackage{siunitx}
\usepackage[nameinlink]{cleveref}
\usepackage[version=3]{mhchem}
\usepackage[breaklinks]{hyperref}
\hypersetup{allcolors=blue, colorlinks=true}

\providecommand{\mysc}[1]{#1}

\usepackage{my_acronyms}

\graphicspath{{figures/}}

\newcommand{\ua}{\uparrow}    
\newcommand{\ub}{\downarrow}  

\providecommand\ket[1]{\lvert#1\rangle}

\providecommand{\I}{\mathrm{i}}
\providecommand\pdiff[2]{\frac{\partial #1}{\partial #2}}
\providecommand\op[1]{\hat{#1}}
\providecommand{\vect}[1]{\vec{#1}}
\renewcommand{\d}{\mathrm{d}}

\begin{document}

\title{Shock Waves in a Superfluid with Higher-Order Dispersion}

\author{M.~E.~Mossman}
\affiliation{Department of Physics and Astronomy, Washington State University, Pullman, WA, USA 99164}

\author{E.~S. Delikatny}
\affiliation{Department of Physics and Astronomy, Washington State University, Pullman, WA, USA 99164}

\author{Michael McNeil Forbes}
\email{m.forbes@wsu.edu}
\affiliation{Department of Physics and Astronomy, Washington State University, Pullman, WA, USA 99164}
\affiliation{Department of Physics, University of Washington, Seattle, WA 98105, USA}

\author{P.~Engels}
\email{engels@wsu.edu}
\affiliation{Department of Physics and Astronomy, Washington State University, Pullman, WA, USA 99164}

\preprint{INT-PUB-20-013}

\begin{abstract}
  \noindent
Higher-order dispersion can lead to intriguing dynamics that are becoming a focus of modern hydrodynamics research.
  Such systems occur naturally, for example in shallow water waves and nonlinear optics, for which several types of novel dispersive shocks structures have been identified.
  Here we introduce ultracold atoms as a tunable quantum simulations platform for higher-order systems.
  Degenerate quantum gases are well controlled model systems for the experimental study of dispersive hydrodynamics in superfluids and have been used to investigate phenomena such as vortices, solitons, dispersive shock waves and quantum turbulence.
  With the advent of Raman-induced spin-orbit coupling, the dispersion of a dilute gas Bose-Einstein condensate can be modified in a flexible way, allowing for detailed investigations of higher-order dispersion dynamics.
  Here we present a combined experimental and theoretical study of shock structures generated in such a system.
  The breaking of Galilean invariance by the spin-orbit coupling allows two different types of shock structures to emerge simultaneously in a single system.
  Numerical simulations suggest that the behavior of these shock structures is affected by interactions with vortices in a manner reminiscent of emerging viscous hydrodynamics due to an underlying quantum turbulence in the system.
  This result suggests that spin-orbit coupling can be used as a powerful means to tun the effective viscosity in cold-atom experiments serving as quantum simulators of turbulent hydrodynamics, with applications from condensed matter and optics to quantum simulations of neutron stars.
\end{abstract}

\maketitle

\glsreset{SOC}
\noindent The dynamics of systems with higher-order dispersion is currently at the forefront of modern hydrodynamics research.
While systems with parabolic dispersion are well understood, higher-order corrections lead to intriguing and peculiar effects that are relevant for systems including shallow water waves and optical media~\cite{Mohamadou2010,Malaguti2014,El2016_2,Liu2018,Zhao2020}.
For example, the dynamics of shock waves in systems with higher-order dispersion have recently been investigated in Sprenger et al.~\cite{Sprenger:2017} using a fifth-order Korteweg-deVries equation to describe classical shallow water waves.
Due to the complexity of the dynamics, many open questions remain, and the fundamental nature of these shock waves is only starting to be explored.

Superfluids, such as dilute-gas \glspl{BEC}, provide a powerful platform for studying dispersive dynamics.
By immersing a dilute-gas \gls{BEC} into an appropriately tuned laser field generated by Raman beams, one can induce \gls{SOC} in the \gls{BEC}~\cite{Lin2009}.
This modifies the single-particle dispersion from parabolic, $E(p) = p^2\!/2m$, to a double-well structure with higher-order terms, similar in form to band structures found in condensed matter systems.
Features of the dispersion can be tailored in experiments: for example, by changing the intensity of the Raman beams, one can manipulate the curvature of the dispersion.
\begin{figure}[t!]
  \includegraphics[width=\columnwidth]{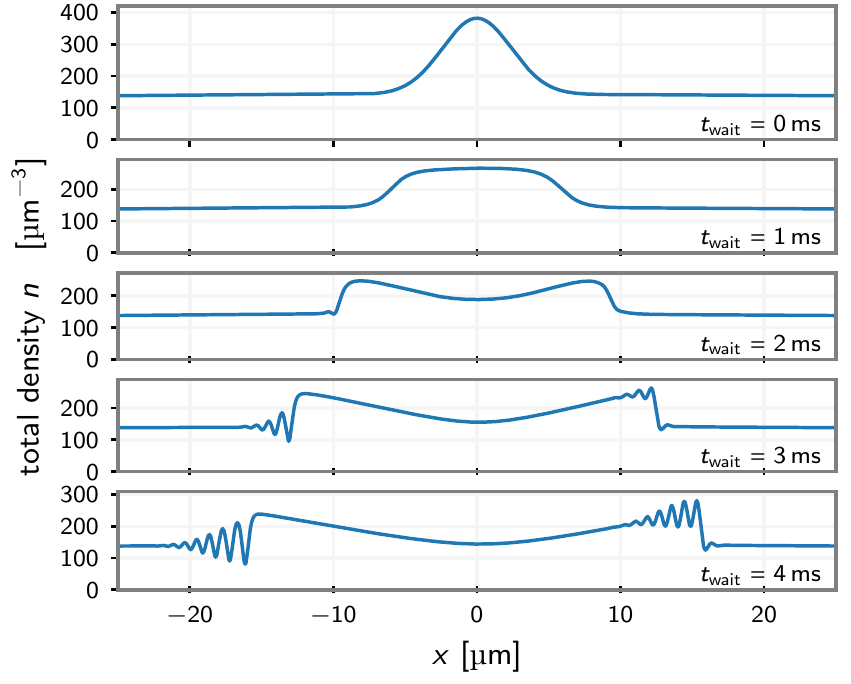}
  \caption{\label{fig:DSW}%
    \textbf{1D numerical simulation modelling the development of dispersive shock waves in a \gls{SOC} \gls{BEC}.}
    At $t_\mathrm{wait}=\SI{0}{ms}$, an initial density perturbation is formed by an attractive Gaussian potential at the center of the system.
    The potential is suddenly switched off at $t_\mathrm{wait}=\SI{0}{ms}$ and the initial density perturbation spreads outwards ($t_\mathrm{wait}=\SI{1}{ms}$), forming two travelling peaks ($t_\mathrm{wait}=\SI{2}{ms}$) and developing into \glspl{DSW} moving in opposite directions ($t_\mathrm{wait}=\text{\SIlist{3;4}{ms}}$).
    The structure of a \gls{DSW} is highly dependent on the background dispersion. 
    In a \gls{SOC} \gls{BEC}, the breaking of Galilean invariance induces two distinct \gls{DSW} structures in a single system.
    To the right, the solitary wavetrain lags behind the solitary wave edge (shock front), while to the left, the solitary wavetrain travels faster than the large amplitude shock front.
}
\end{figure}
\exclude{In a recent experiment, a \gls{BEC} with equal Rashba and Dresselhaus \gls{SOC} was shown to exhibit features consistent with negative-mass hydrodynamics~\cite{Khamehchi:2017}.
The induced negative curvature of the dispersion leads to regions of negative effective inertial mass where particles accelerate in the direction opposite to any applied force, strongly affecting the dynamics of the \gls{BEC}.}

\glsreset{DSW}
A \gls{BEC} with \gls{SOC} constitutes an exotic medium through which topological defects, phonons, and shock waves can propagate. 
These features have characteristics that are strongly correlated with the properties of the underlying medium. 
For instance, in a conventional \gls{BEC}, small-amplitude phonons propagate near the speed of sound in the medium at long wavelengths~\cite{Bogoliubov1947, Andrews1997}, and as the wavelength decreases, the propagation speed increases slightly~\cite{Chang2008}.
However, in a \gls{SOC} \gls{BEC}, the dispersion can be modified so that short wavelength modes travel more slowly in specific directions.
This has a profound impact on the shape of \glspl{DSW} that develop from non-linear interactions in the system~\cite{Hoefer2006}. 
A prototypical example of this is demonstrated in \cref{fig:DSW} showing the results of one-dimensional numerical simulations using realistic \gls{SOC} parameters to form the single component dispersion, shown in \cref{fig:fig6}.  

Shock waves generated in a superfluid medium are typically considered to be dispersive: instead of becoming infinitely steep,
the shock front is smoothed by gradients in the kinetic energy (dispersion).
In direct contrast to this, classical shock waves are smoothed by dissipative effects such as viscosity.
Depending on the amplitude of the excitations and the geometry of the system, shock waves in a superfluid can decay into a variety of intricate structures determined by the dimensionality of the system~\cite{Kevrekidis2015}, in-part due to the presence of snaking instabilities along confined transverse directions~\cite{Kuznetsov:1986, Muryshev2002, Brand:2002, Hoefer:2016}.
In one-dimensional systems, effectively realized in elongated trap geometries with tight radial confinement, superfluid shock waves remain dispersive~\cite{Chang2008, Hoefer2009, Meppelink2009}.
As the dynamics probe additional dimensions, however, shock waves can appear to be dissipative, despite a lack of dissipation in the superfluid systems~\cite{Joseph2011,Mossman2018}.
This effective viscosity arises from the generation of quantized superfluid vortices through snaking instabilities, resulting in a turbulent fluid that can be modelled by one-dimensional \gls{VSW} theory.
We note that viscosity can appear in superfluids due to interactions with the normal component (mutual friction) and as intrinsic shear viscosity, but these effects are much smaller than those discussed here which can be reproduced with purely conservative simulations.

In this work, we showcase an example of the rich dynamics supported by superfluids with a higher-order dispersion. 
We experimentally demonstrate, with numerical verification, how \glspl{DSW} evolve in a \gls{3D} \gls{SOC} environment, resulting in asymmetric topological features displaying varying amounts of dissipation through viscous-like effects.
Comparing with simulations, we interpret this viscous-like behavior as a manifestation of quantum turbulence arising from interactions of the shock front with vortex rings and solitons.
The dynamics of these defects are subtly affected by the dispersion, resulting in qualitatively different macroscopic behavior.
In this way, \gls{SOC} provides a way to tune the effective viscosity of the macroscopic hydrodynamics realized in turbulent quantum fluids.
This provides experimental control over cold-atom systems used as quantum simulators of turbulent hydrodynamics.

\section{Results}
\subsection{Experimental Setup}
\begin{figure}[t]
  \includegraphics[width=\columnwidth]{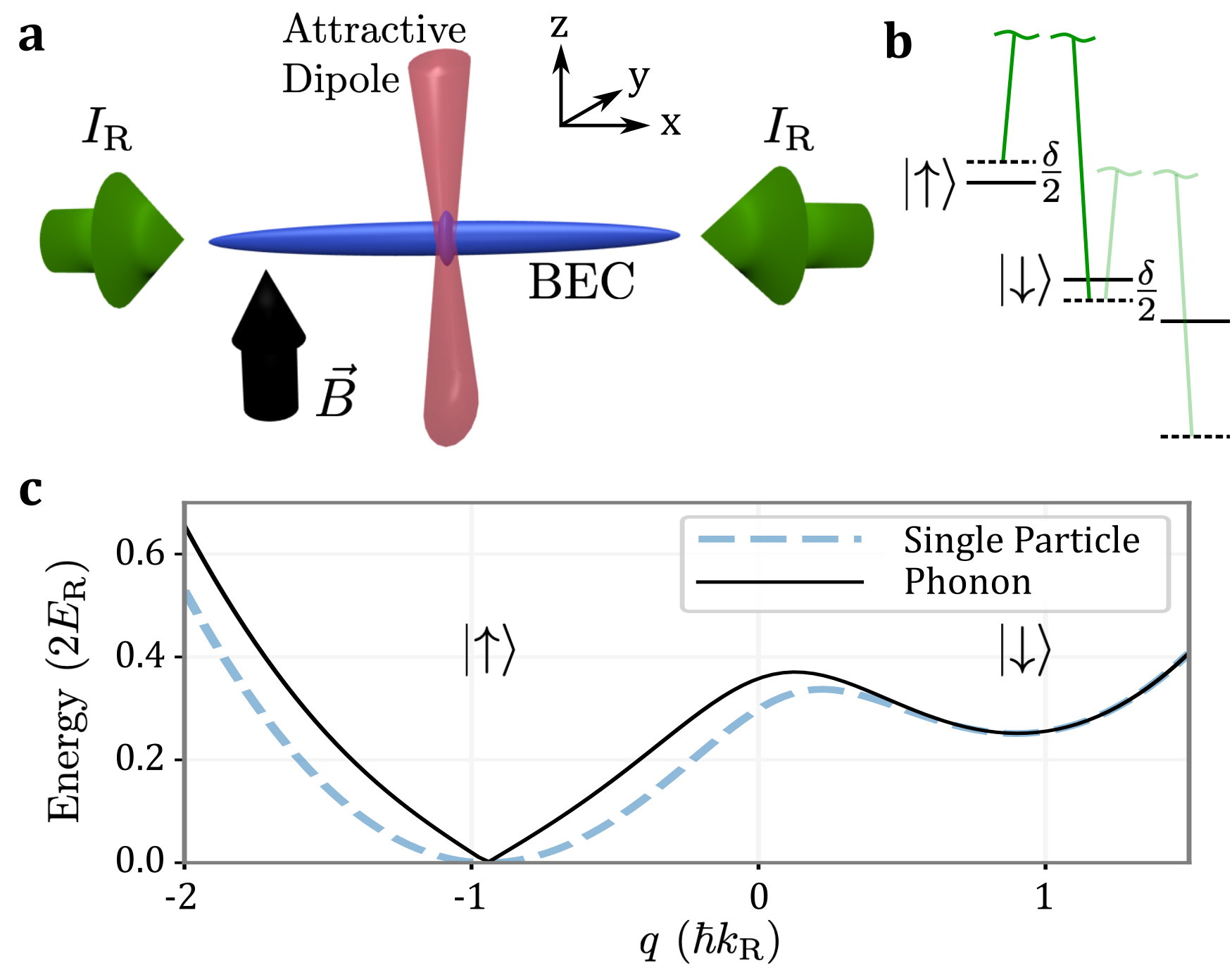}
  \caption{\label{fig:1}%
    \textbf{Experimental setup, level scheme and dispersion.}
    \textbf{a} Experimental setup: an elongated \gls{BEC} is prepared with spin-orbit coupling in the presence of a vertical magnetic field with two counter propagating Raman beams (green).
    An attractive dipole sheet (red) is applied at the center of the \gls{BEC}. 
    \textbf{b} Relevant level scheme of the \ce{^{87}Rb} $\ket{F=1}$ hyperfine states. The energies of the levels are shifted due to the Zeeman effect.
    Raman beams couple the $\ket{\ua}$ and $\ket{\ub}$ pseudo-spin states.
    \textbf{c} Two component excitation spectrum (black solid) and single particle dispersion (blue dashed) for the experimental parameters $\hbar\Omega = 1.5 E_\mathrm{R}$ and $\hbar\delta = 0.54 E_\mathrm{R}$.
    The blue-shaded area indicates quasimomenta with negative effective mass.
    The phonon dispersion has been shifted in the plot along the quasimomentum axis to line up with the single particle dispersion minimum for convenience.
  }
\end{figure}

To investigate the excitation dynamics, we employ an elongated \gls{BEC} of \smash{\ce{^{87}Rb}} atoms, confined in an optical crossed-dipole trap [See Methods for detailed experimental parameters].
The \gls{BEC} is cigar shaped with an aspect ratio of approximately 80:1, and the long axis of the \gls{BEC} is oriented horizontally as shown in \cref{fig:1}a.
A uniform bias field in the $z$-direction splits the $F=1$ hyperfine ground state in accordance to the Zeeman shift.

Spin-orbit coupling, with its associated double-well dispersion, is induced by applying two counter-propagating Raman beams that couple the $\ket{F,m_F} = \ket{1,-1}$ and $\ket{1,0}$ state, which we designate as two spin orientations $\ket{\ua}$ and $\ket{\ub}$ of a pseudo-spin $1/2$ system, respectively [see \cref{fig:1}b and Methods].
The height of the central hump in the single-particle dispersion (near quasimomentum $q=0$ [see \cref{fig:1}c]) depends on  the Raman coupling strength $\Omega$, which can be adjusted in the experiment by the intensity of the Raman beams, $I_\mathrm{R}$.
The energetic offset of the two local minima of the dispersion depends on the detuning $\delta$ of the Raman coupling, which can be set by the frequency difference between the two Raman beams. 
The experimentally realized single particle dispersion and the associated two-component phonon dispersion are shown in \cref{fig:1}c by the dashed blue and solid black lines, respectively.
See \cref{supp:1} for more information. 
Energies and momenta are measured in units of the recoil energy, $E_{\mathrm{R}} = \hbar^2 k_{\mathrm{R}}^2/2m$, and recoil momentum, $k_{\mathrm{R}} = 2\pi/\lambda_{\mathrm{R}}$, where $\lambda_{\mathrm{R}}$ is the Raman laser wavelength.
The \gls{BEC} is prepared with \gls{SOC} such that the majority amplitude of atoms are in the $\ket{\ua}$ spin state. 
The direction of \gls{SOC} coincides with the long axis of the \gls{BEC} such that the direction of positive quasimomentum $+q$ is in the $+x$-direction, as indicated in \cref{fig:1}a. 

\begin{figure}[htb]
  \includegraphics[width=\columnwidth]{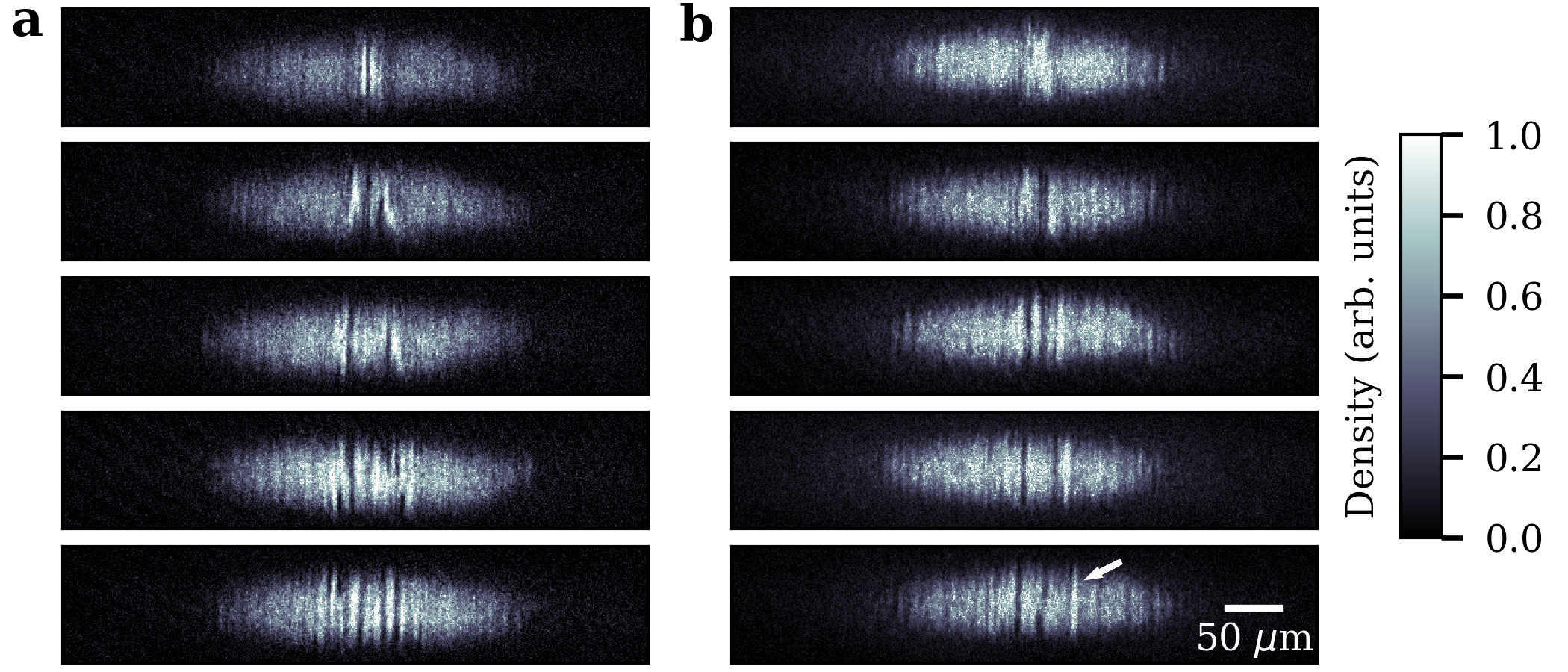}
  \caption{\label{fig:expt_fast}%
    \textbf{Experimental results with and without \gls{SOC}.}
    Absorption images acquired after a \SI{10.1}{ms} time-of-flight expansion show the dynamics of outward moving excitations in a \gls{BEC} prepared \textbf{a} without and \textbf{b} with \gls{SOC} at times $t =  \text{\SIlist{0;4;8;12;16}{ms}}$ after a $U_\mathrm{b} = \SI{-90}{nK} \approx -1.6\mu$ attractive potential has been applied for \SI{10}{ms}.
   A Stern-Gerlach technique is used during time-of-flight to vertically separate the spin states in \textbf{b}, where only the majority component is shown. The arrow in the last panel of \textbf{b} indicates a highly reproducible peak propagating to the right that is discussed in the main text.
}
\end{figure}

An additional dipole sheet aligned perpendicular to the long axis of the \gls{BEC} creates an attractive Gaussian potential for the atoms at the center of the \gls{BEC}. 
This vertical dipole sheet is pulsed on for \SI{10}{ms} after the system has been prepared with \gls{SOC}, resulting in excitations that propagate outwards along the along axis towards the edges of the \gls{BEC}. 
The depth of this dipole potential $U_{\mathrm{b}}$ can be varied to generate large or small initial excitations in the \gls{BEC}.
In this work, $U_\mathrm{b}$ is on the order of the chemical potential of the majority component state ($\ket{\ua}$) in the \gls{SOC} \gls{BEC}, $\mu$.
To analyze the dynamics, absorption imaging is performed after a \SI{10.1}{ms} time-of-flight expansion. A Stern-Gerlach technique is used to vertically separate the spin states during the imaging procedure. 
Representative images obtained this way are presented in \cref{fig:expt_fast}, where the $\ket{\ub}$ state is not shown due to very low number atoms in the minority component.

\begin{figure*}
  \includegraphics[width=\textwidth]{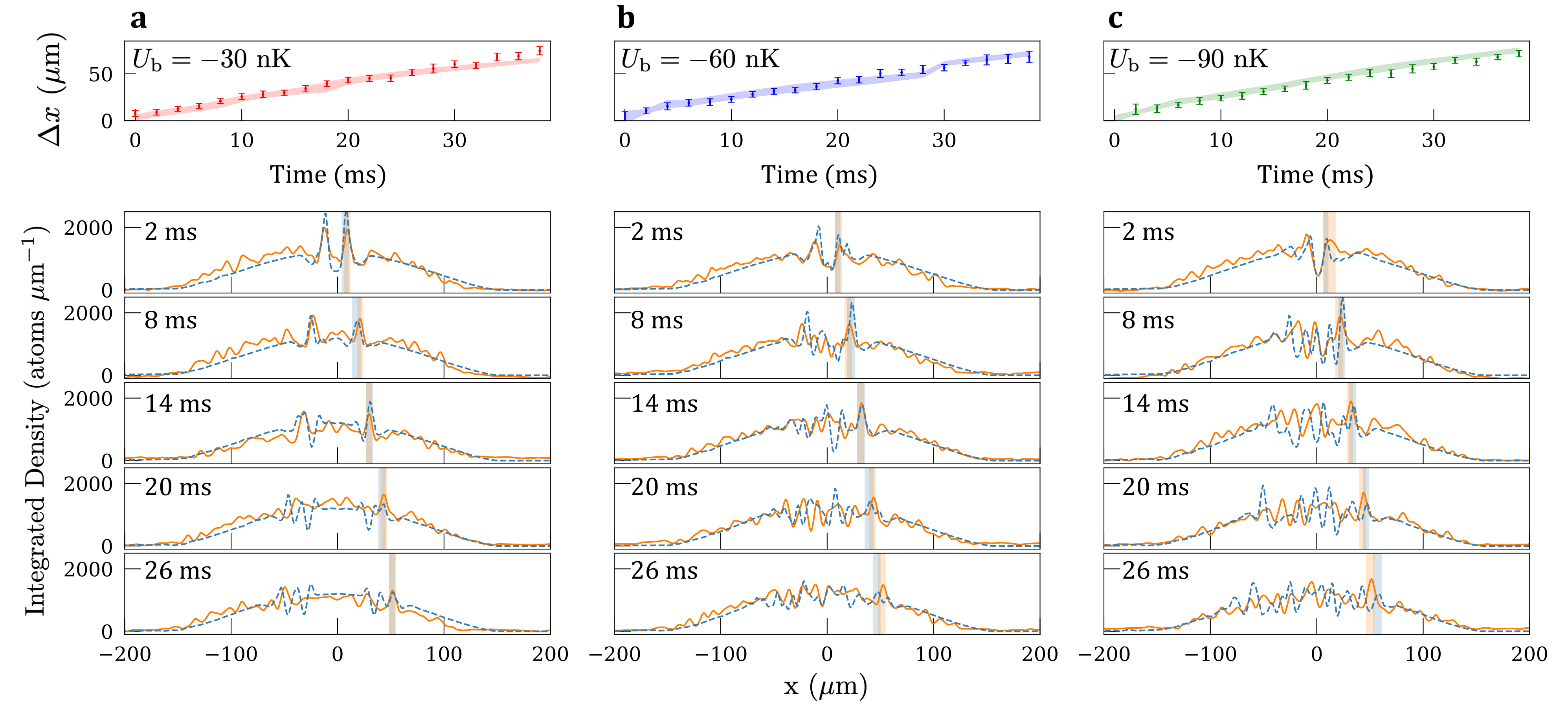}
  \caption{\label{fig:comparison}%
    \textbf{Experimental results and numerical simulations.}
    Analysis and integrated cross sections of an expanded \gls{BEC} prepared with \gls{SOC} after an attractive potential with \textbf{a} $U_\mathrm{b} = \SI{-30}{nK}$, \textbf{b} $U_\mathrm{b} = \SI{-60}{nK}$, or \textbf{c} $U_\mathrm{b} = \SI{-90}{nK}$ has been pulsed on at the center of the cloud for \SI{10}{ms}.
    After the potential is switched off, the system is allowed to evolve for a time $t_\mathrm{wait}$ in the presence of \gls{SOC} prior to \SI{10.1}{ms} time-of-flight expansion.
    The top row shows the position of the right moving excitation over time for \gls{GPE} simulations (shaded region) and experiment (data points) for each potential depth, measured using a Gaussian fit function.
    The error bars and bands for both experiment and numerical results show the $2\sigma$ waist of the fitted Gaussian to the measured excitation.
    Quantitative results for the experiment are stated in the main text. 
    In the lower panels, integrated cross sections for both \gls{GPE} simulation (blue) and experiment (orange) are provided for each potential depth after $t_\mathrm{wait} = \text{\SIlist{2;8;14;20;26}{ms}}$, where the shaded regions reflect the data presented in the top row.
    The \gls{GPE} simulations presented here are performed using axial symmetry, which forces topological defects to align along the imaging axis.
    For more information, see \cref{supp:1}.
    The simulations have been modified to reflect the optical resolution of the experiment ($\sim\SI{2}{\micro m}$) using a Gaussian convolution method.}
\end{figure*}

\subsection{Experimental Results}
We have performed a systematic study of the dynamics following the sudden switch-off of the dipole sheet as a function of the evolution time $t_{\mathrm{wait}}$ and initial potential strength $U_{\mathrm{b}}$. 
A synopsis is presented in \cref{fig:comparison} and reveals the following features, the interpretation of which is confirmed by our matching numerical simulations:
In the absence of \gls{SOC}, the left-travelling and right-travelling excitations qualitatively behave the same as they propagate to the edges of the \gls{BEC}, forming vortex rings and dark solitons.
When strong \gls{SOC} is applied to the system, parity is broken and an asymmetric behavior is observed between the two directions.
This asymmetric behavior is highly dependent on the depth of the initial potential with respect to $\mu$ and on the coupling strength of the \gls{SOC}.
For a system where the \gls{SOC} coupling strength $\Omega$ and detuning $\delta$ are fixed, the following behavior is found:

\begin{enumerate}
\item When $U_\mathrm{b} < \mu$, excitations move outwards from the center of the \gls{BEC}, displaying no discernible difference between the cases with and without \gls{SOC}. 
\item When $U_\mathrm{b} \gtrsim \mu$, the excitation propagating to the right consistently forms a well-defined peak that becomes particularly pronounced during the expansion dynamics and travels outward from the center towards the right edge of the cloud at a relatively constant velocity. For clarity, this peak is indicated with a white arrow in the lower right image of \cref{fig:expt_fast}.
Quantitative analysis of the right-travelling excitation yield experimental speeds of $1.63 \pm 0.04$~mm/s for $U_\mathrm{b}=$\SI{-30}{nK}, $1.64 \pm 0.05$~mm/s for $U_\mathrm{b}=$\SI{-60}{nK}, and $1.68 \pm 0.06$~mm/s for $U_\mathrm{b}=$\SI{-90}{nK}.
This excitation is highly reproducible and observed to have a lifetime comparable to small-amplitude excitations in past phonon excitation experiments~\cite{Andrews1997, Chang2008, Meppelink2009}. 
In addition, solitonic excitations are seen in the experimental images, and numerical simulations of the \gls{GPE} identify the generation of a collection of defects, including solitons, solitonic vortices, and vortex rings, during the \SI{10}{ms} pulse of the attractive potential.
The positions of these features depend subtly on small details, such as a tiny tilt in the dipole sheet, which are expected to vary in the experiment from shot to shot.
\end{enumerate}

\subsection{Numerical Results}
To understand the experimental results, numerical simulations of a coupled set of \glspl{GPE} are performed:
\begin{subequations}\label{eq:H2}
  \begin{gather}
    \I\hbar\pdiff{}{t}
    \begin{pmatrix}
      \ket{\ua}\\
      \ket{\ub}
    \end{pmatrix}
 =
   \begin{pmatrix}
     \frac{\op{p}^2}{2m} + V_\ua & \frac{\Omega}{2}e^{2\I k_R x}\\
     \frac{\Omega}{2}e^{-2\I k_R x} & \frac{\op{p}^2}{2m} + V_\ub
   \end{pmatrix}\cdot
   \begin{pmatrix}
     \ket{\ua}\\
     \ket{\ub}
   \end{pmatrix},\\
    V_{\ua/\ub} = -\mu \pm \frac{\delta}{2} + g_{\ua\ua/\ua\ub}n_\ua + g_{\ua\ub/\ub\ub}n_\ub
  \end{gather}
\end{subequations}
where $\op{p} = -\I\hbar\vect{\nabla}$ is the momentum operator, $\mu$ is the chemical potential in the \gls{SOC} system, $g_{ab} = 4\pi\hbar^2 a_{ab}/m$, and $a_{ab}$ are the $s$-wave scattering lengths (with a, b = $\ua$ or $\ub$).
For \smash{\ce{^{87}Rb}}, $a_{\ua\ua}= 100.40a_0$, $a_{\ub\ub} = 100.86a_0$, and $a_{\ua\ub}=100.41a_0$ where $a_0$ is the Bohr radius.
The system is prepared in the ground state with a \gls{TF} cloud radius of $x_{\mathrm{TF}} = \SI{150}{\micro\meter}$ along the long axis, corresponding to $N_\ua=\num{206000}$ and $N_\ub=\num{6000}$ atoms in the condensate. The \gls{SOC} parameters are $\hbar\Omega = 1.5 E_\mathrm{R}$ and $\hbar\delta = 0.54 E_\mathrm{R}$.
To reduce computational costs, cylindrical symmetry is employed about the long axis of the trap.
The system is evolved in real time following the experimental protocol including the imaging procedure, which we implement in an expanding coordinates system as discussed in~\cite{Castin:1996}.
See \cref{supp:2} for details.
This introduces some significant artifacts by restricting vortices to be vortex rings, but allows us to fully simulate the experimental procedure including the expansion and imaging.
Limited full \gls{3D} simulations of the in situ dynamics confirming the behavior discussed here are shown in \cref{fig:states}.

\subsection{Interpretation of Results}
\begin{figure*}[bt]
  \includegraphics[width=\textwidth]{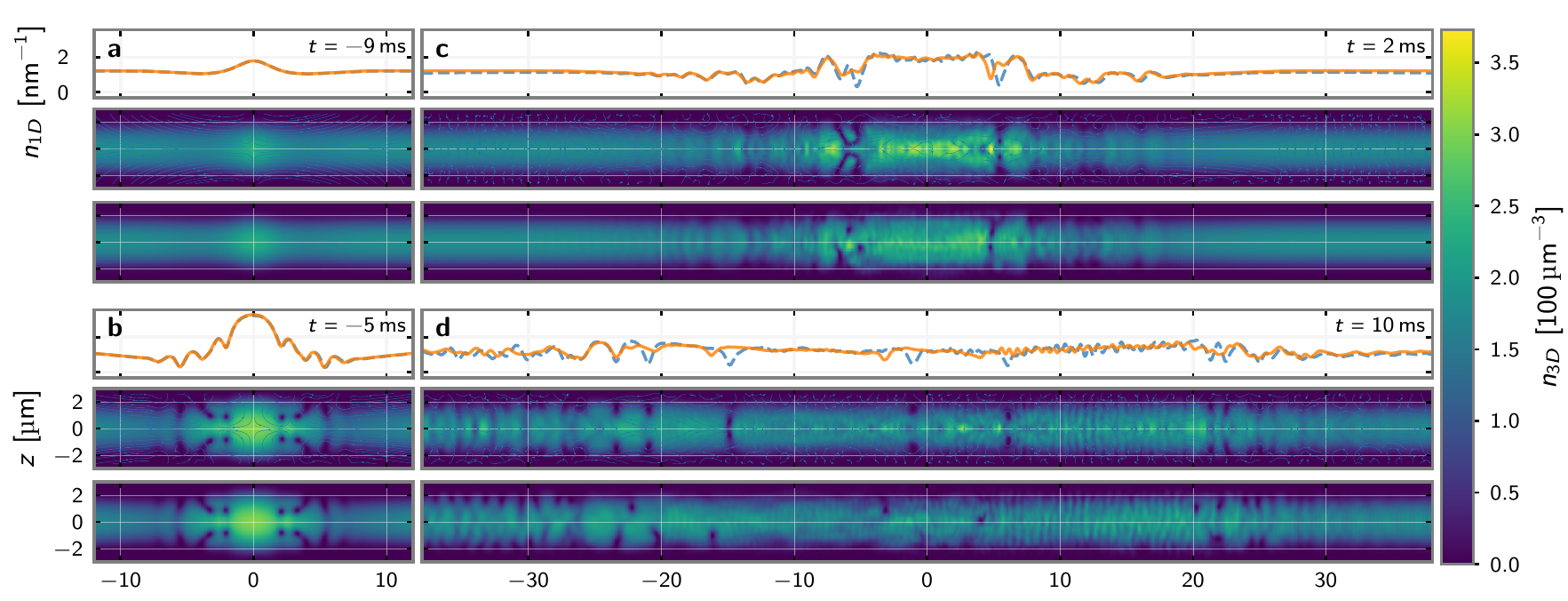}
  \caption{\label{fig:states}%
  \textbf{Numerical simulations with and without barrier tilt.}
    Simulated in situ images (prior to the expansion) of the four distinct stages of evolution from a simulation of the experiment with a \SI{-60}{nK} attractive potential.
    Here we compare axially symmetric simulations to 3D simulations with a small 1\% y-tilted Gaussian dipole beam $V_{DB} \propto \exp{\bigl((x+0.01y)^2/2\sigma\bigr)}$.
    Upper frames are the integrated line-density $n_{1D}(x) = \int \d{y}\d{z}\; n(x,y,z)$, with axially symmetric data (blue), and \gls{3D} tilted data (orange).
    Middle frames are slices $n_{3D}(x,y) = n(x,y,0)$ for axially symmetric simulations with streamlines of the current density $n_\ua\vect{v}_\ua + n_\ub\vect{v}_\ub$. 
    Bottom frames are slices $n_{3D}(x,y) = n(x,y,0)$ from the 3D tilted simulations. 
    \textbf{a} Initial flow of the \gls{BEC} into the region of the attractive potential.
    \textbf{b} Formation of several vortex rings due to snaking instabilities induced by this flow.
    The pattern of rings here is quite symmetric, even in the tilted case.
    Vortex rings appear as reduced density in the 1D plots with mild dependence on the ring radius.
    \textbf{c} Formation of outgoing \glspl{DSW} after the attractive potential is removed at $t=\SI{10}{ms}$.
    In this frame, the \glspl{DSW} are just starting to interacting with the first seeded vortex ring: without the seeding rings, the structure of these \glspl{DSW} is similar to that shown in \cref{fig:DSW}.
    In the tilted case, the vortex rings rapidly decay into vortices that break the axial symmetry and average to significantly smoother integrated line-density.
    \textbf{d} Results of \glspl{DSW} interacting with initial vortex rings.
    Note that many fully-formed and stable vortices exist on the left, while fewer vortices survive on the right. 
    This is attributed to an increased number of vortex annihilation events to the right, evident by the manifestation of short wavelength oscillations in the integrated cross sections.
    Some features are sensitive to the tilt, such as the vortices near $x=0$ whose location shifts by several microns in the integrated 1D density.
    Others remain robust, such as the right-moving shock wave and corresponding peak at $\text{\SIrange{20}{25}{\micro m}}$.
  }
\end{figure*}

As demonstrated in the experimental absorption images \cref{fig:expt_fast}, a striking effect of the modified dispersion is the apparent stabilization of the right-moving shock wave, leading to a highly reproducible peak seen in the expansion images that is traveling to the right.
This feature is reproduced by our numerical simulations, allowing us to probe the microscopic mechanism for this stabilization.
Our numerics are summarized in \cref{fig:states}. 
Details and animations can be found in \cref{supp:3}.

The numerics show the following progression of events.
Shortly after turning on the attractive potential, fluid is drawn into the central region of the trap where the potential is located (\cref{fig:states}a).
The subsequent flow induces a snaking instability~\cite{Kuznetsov:1986, Muryshev2002, Brand:2002, Hoefer:2016} seen in \cref{fig:states}b, forming vortex rings on either side of the growing central excitation, or bulge.
During the initial stages of evolution, vortex rings form quite symmetrically on both sides. 
Most have their central flow oriented towards the center of the cloud, however, with increasing barrier strength some vortex rings form with central flow facing outward.

After the attractive potential is turned off, the central bulge expands along the axis of the trap, as shown in \cref{fig:states}c.
This can be described by decomposing the bulge as a superposition of left- and right-moving bulges (phonons), which move outward at approximately the local speed of sound once the attractive potential is suddenly switched off.
Due to the non-linear interaction, these left- and right-moving bulges quickly form \glspl{DSW}, the orientation and polarization~\cite{El:2016} of which are sensitive to the curvature of the dispersion as shown in \cref{fig:DSW}.
In particular, the left-moving \gls{DSW} forms a leading soliton train as short-wavelength components travel faster than the solitary wave edge of the bulge. 
On the right, a small-wavelength soliton train trails behind the bulge.
This has a simple explanation in terms of the modified dispersion.
On the right, the phonon dispersion has negative curvature, and both group and phase velocities
of the short-wavelength modes are \emph{slower}~\cite{El:2016, Sprenger:2017}.
Discerning these features in an experiment would require high-resolution in-situ imaging as they are on the order of the healing length and they do not survive the expansion imaging procedure.

As these outward traveling shock waves overtake the initially seeded vortex rings, intriguing dynamics ensue and a complex interaction develops between the rings and the shock front, as shown in \cref{fig:states}d.
In particular, the vortices absorb energy and momentum from the shock front, causing the shock to dissipate as if it were a viscous shock wave (\gls{VSW}) even though the total energy is conserved by the system.
This is consistent with previous observations of \glspl{VSW} in superfluids as a result of dimensional reduction~\cite{Joseph2011, Mossman2018} where a dissipationless superfluid in \gls{3D} is described by viscous hydrodynamics in \gls{1D} after integrating over the transverse directions.
The modified dispersion plays an important role here, significantly suppressing these dissipative effects.
In \cref{fig:states}d one can see a large number of vortex rings on the left side of the cloud, while very few remain on the right.
Examining the detailed dynamics (see the \cref{supp:3}), we see that vortex-vortex and vortex-shock front interactions are more likely to trigger vortex annihilation on the right side of the cloud than on the left.
As a result, fewer vortices remain on the right and less energy is dissipated from the shock front, leading to less effective viscosity and to the stabilization effect we observe in experiments with modified dispersion.

One might wonder if the asymmetry is due to the initial asymmetric form of the \glspl{DSW} shown in \cref{fig:DSW}, in particular noting that the strong leading soliton train on the left might trigger the formation of more vortices.
While this likely plays a role, it appears that the vortices seeded in \cref{fig:states}d, \cref{fig:states}b are crucial to the observed dynamics, at least at these potential strengths.
Both numerics and experiment reveal that using a shallower potential, such as $U_\mathrm{b} = \SI{-30}{nK}$ (see \cref{fig:comparison}a), decreases the number of vortices generated initially in the system, thus greatly reducing the effective viscosity, and shocks propagate in both directions without significant dissipation.

Explaining the exact microscopic mechanism for the enhanced likelihood of vortex annihilation with modified dispersion requires further investigation, but we anticipate that this is largely due to the presence of a modulational instability in the region of negative effective mass (shaded region in \cref{fig:1}c)~\cite{Khamehchi:2017}.
As the shock front passes through a vortex ring, it can induce portions of its flow to enter this region where modulational instabilities can manifest.
Our numerics reveal that this often triggers the vortex ring to rapidly collapse or expand out of the system, effectively decaying to many high-frequency phonons seen as rapid, near-stationery oscillations on top of the simulations in \cref{fig:states}d.
In contrast, vortex interactions with the shock front moving to the left change the diameter of the vortex rings, triggering fewer annihilation events, and leaving them free to absorb the energy from the passing shock wave.
Annihilations from the modulational instability occur primarily in the center of vortex rings when the relative flow from the passing shock front increases the quasimomentum into the negative mass region.
Notably, the rapid oscillations from the modulation instability are also seen developing on the cusp of the outward traveling right side \gls{DSW} in the raw $U_b = \text{\SIlist{-30;-60;-90}{nK}}$ numerical data sets. 

\subsection{Expansion Dynamics}
Structures induced by \gls{SOC} and topological defects formed during \gls{DSW} decay have length scales on the order of a healing length.  
These length scales are below the imaging resolution in our experimental setup. 
Therefore, time-of-flight imaging with \SI{10.1}{ms} expansion time is used.
During this expansion, features like solitons and vortices widen and thus can be resolved by the imaging system~\cite{Anderson2001}.
We have performed numerical simulations of the expansion dynamics which reveal that
this process is nontrivial and the structures of the excitations change considerably during this time.
We find that the expansion process significantly enhances the peak of the \glspl{DSW}, allowing it to be clearly imaged after the experiment:
After the \gls{BEC} is released from the trap, the gas expands rapidly in the radial direction, reducing the density by more than a factor of $10$ in \SI{2}{ms}, and rendering the gas essentially non-interacting. 
In the remaining \SI{8}{ms} of expansion, the various frequency components determined from the bare particle dispersion separate with velocity $v = \hbar k/m$, where $k$ is the wave vector of the frequency component. 
What remains is a highly enhanced peak moving with the characteristic momentum of the shock wave.
See \cref{supp:4} for more information and animations.

\section{Discussion}
Using an attractive dipole sheet to generate large amplitude excitations on the background of a \gls{SOC} \gls{BEC}, we are able to probe the effects of higher-order dispersion on the non-linear dynamics in an ultracold atomic system.
The experimental results show a clear asymmetry in the non-linear dynamics in the presence of \gls{SOC}, manifesting an enhanced stability of shock fronts propagating into the direction of higher-order dispersion, in agreement with \gls{GPE} simulations.
Within the numerical simulations, one is able to resolve the microscopic origin of this stability: the \gls{SOC} significantly modifies the dynamics and stability of vortices in the region of modified dispersion, reducing their ability to dissipate energy from the shock wave.
While it has been shown that the presence of \gls{SOC} significantly alters the structure of vortices~\cite{Radic:2011}, no comparable study of the effect on their dynamics has been performed.

The left-moving shock front decays rapidly, leaving behind a wake of vortices, while the right-moving shock front remains quite stable.
We interpret the observed asymmetry as a manifestation of quantum turbulence:
Viewed in terms of \gls{1D} \gls{VSW} theory, the vorticity induced in the system provides a mechanism to absorb energy, resulting in an effective viscosity in the \gls{1D} theory, similar to that seen in previous superfluid experiments~\cite{Joseph2011, Mossman2018}.
This effect is qualitatively consistent with our results, but further analysis is required to quantify the effective viscosity.
In this language, the modified dispersion here significantly alters the vortex dynamics in comparison to previous cited works, reducing the effective viscosity for the right-moving shock front.
Thus, \gls{SOC} provides an effective tool for modifying the underlying dynamics of vortices, and thereby tuning the effective viscosity of the long-range hydrodynamic effective theory.
Such control is essential for using cold-atoms as effective quantum simulators for turbulent fluid dynamics.

\exclude{
\subsection{Code and Data Availability}
All relevant code used for numerical studies and datasets appearing in this work are available from the corresponding authors on reasonable request.
}

\begin{acknowledgments}
We thank Prof.\@ Mark Hoefer and Patrick Sprenger for thoughtful and in-depth discussions concerning the behavior and shape of shock structures in higher-order dispersions. 
M.E.M.\@ and P.E.\@ are supported by the
\gls{NSF} through Grants No.\@ \mysc{PHY-1607495} and \mysc{PHY-1912540}. 
E.S.D.\@ and M.M.F.\@ are supported by the \gls{NSF} through Grant No.\@ \mysc{PHY-1707691}. 
\end{acknowledgments}

\exclude{
\subsection*{Author contributions} 
M.E.M., P.E., and M.M.F.\@ conceived the experiment;
M.E.M.\@ and P.E.\@ performed the experiments; 
E.S.D.\@ and M.M.F.\@ performed the theoretical calculations; 
P.E.\@ and M.M.F.\@ supervised the project.
All authors discussed the results and contributed to the writing of the manuscript.

\subsection*{Competing financial interests} The authors declare no competing financial interests.
}

\newcommand{\AxialA}{\href{https://youtu.be/q0ItfdDurec}{Axial \SI{30}{nK}}}
\newcommand{\AxialB}{\href{https://youtu.be/DmYjGtJcRKQ}{Axial \SI{60}{nK}}}
\newcommand{\AxialC}{\href{https://youtu.be/ApzQK4d8dtg}{Axial \SI{90}{nK}}}

\appendix
\section{Movies}
The following are links to movies on YouTube showing various results:
\begin{description}
\item[\AxialA, \AxialB, \AxialC]    
Axially symmetric simulations of the data in ~\cref{fig:comparison}, before \gls{ToF} expansion.
\item[\href{https://youtu.be/YO6fEDD-9zI}{Axial \SI{90}{nK} ToF twait \SI{26}{ms}}]
Axial simulation of 10.1ms \gls{ToF} expansion starting after \SI{26}{ms} in trap evolution.
\item[\href{https://youtu.be/UcXwBZ7liJE}{Axial 3D tilt \SI{30}{nK}}]: 
A comparison between the Axial, tube, 3D, and 1\% tilt 3D numerical methods at barrier depth \SI{-30}{nK}.
\end{description}

\section{Experimental Methods and Parameters}
Our experiments are conducted with elongated \glspl{BEC} of \ce{^{87}Rb} atoms. The atoms are confined in an optical crossed-dipole trap with trap frequencies $\{\omega_x, \omega_y,\omega_z \}=2\pi\times\{3.49, 278, 278\}$~\si{Hz}, where the weakly confining direction is oriented horizontally.
A \SI{10}{G} uniform bias field leads to a Zeeman splitting of the hyperfine states. The $\ket{1,-1}$ and $\ket{1,0}$ state are coupled through a two-photon Raman transition, while the $\ket{1,+1}$ state is essentially uncoupled due to the quadratic Zeeman effect. 
After loading into \gls{SOC}, there are approximately \num{2e5} atoms in the majority ($\ket{\ua}$) component of the condensate.
During the experiment, the Rabi coupling strength is $\hbar\Omega = 1.5E_{\mathrm{R}}$.
The detuning of the Raman drive is set to $\hbar\delta = \num{0.54+-0.01} E_{\mathrm{R}} \equiv \SI{2000+-50}{Hz}$, where the uncertainty is given by the stability of the external bias field.

During the preparation and course of the experiment, heating caused by the Raman beams will decrease the condensate fraction, reducing the \gls{1D} longitudinal speed of sound and the equivalent non-\gls{SOC} chemical potential in the majority component spin state from their initial values of $c_{s0}\approx \SI{2.2}{mm/s}$ and $\mu_0\approx \SI{100}{nK}$ to $c_{s}\approx \SI{1.6}{mm/s}$ and $\mu \approx \SI{55}{nK}$, respectively.

An additional vertical dipole sheet, with $\lambda_\mathrm{b} = \SI{850}{nm}$ and Gaussian waists $\{w_x, w_y\}=\{4.8, 27.2\}\,\si{\micro m}$, is focused onto the center of the \gls{BEC}. The extent of the Gaussian profile in the y-direction is larger than the size of the \gls{BEC} in-situ. The beam is pulsed on for \SI{10}{ms} to create excitations at the center of the cloud.
 
 Imaging is performed after \SI{10.1}{ms} time-of-flight expansion during which all laser beams are off, and a Stern-Gerlach technique is used to vertically separate the spin states during the imaging procedure.

\section{GPE Simulations}
To model the experiment, we adjust the chemical potential $\mu$ so that the density of the gas vanishes at $x_{\mathrm{TF}} = \SI{150}{\micro\meter}$ in the \gls{TF} approximation.
These parameters correspond to a lattice spacing of $\d{x} \approx \SI{0.06}{\micro\meter}$ which is sufficiently small compared with the healing length $\xi \approx \SI{0.22}{\micro\meter}$ in the center of the cloud.

We start from the ground state in a harmonic trap with frequencies $\{\omega_x, \omega_y,\omega_z \}=2\pi\times\{3.49, 278, 278 \}$~\si{Hz}.
We then evolve in real time using a 5th-order \gls{ABM} predictor-corrector integration scheme~\cite{Hamming:1973} with step size $\d{t} = \num{6.3}~\mu$s.
We model the dipole sheet with a gaussian potential centered on $x_0$ with a width of $\sigma=\SI{4.8}{\micro m}$.
This potential is turned on and off smoothly using a $C_\infty$ step function over $t_\mathrm{step}=\SI{0.1}{ms}$.
We note that it is important for the accuracy of the \gls{ABM} method that the time-dependent parameters vary smoothly.

To simulate the cloud expansion, we use the scaling procedure described in~\cite{Castin:1996} to scale the radial coordinate without needing to add more lattice points to our simulation.
Since the trapping potential along the cloud is weak, there is very little expansion along the cloud, so we do not scale the coordinate in this direction -- our box is sufficiently large to accommodate this expansion.

Although the dynamics are three-dimensional, the two radial trapping frequencies are approximately equal, and the full \gls{3D} dynamics are well approximated by an axially symmetric geometry.
Axially symmetric simulations can reproduce turbulent features generated by a quantum-mechanical piston in a channel geometry, like that found in Ref.~\cite{Mossman2018}.
Similar agreement between axially symmetric simulations~\cite{Ancilotto:2012} that can reproduce \gls{3D} shock phenomena in channel geometries~\cite{Joseph2011} has also been observed in fermionic superfluids.

There are two differences of note between the experiment and \gls{GPE} simulations.
First, the numerical simulations enforce an axial symmetry, which restricts solitonic excitations, such as vortex rings, to be axially symmetric.
While this is consistent with the experimental geometry, it is well known that small perturbations will destabilize vortex rings, which can evolve relatively quickly into solitonic vortices~\cite{Brand:2002, Becker:2013, Reichl:2013, Ku:2014, Bulgac:2014, Scherpelz:2014, Munoz-Mateo:2014, Ku:2015, Hoefer:2016}.
We have verified by performing unrestricted \gls{3D} simulations (see the \cref{supp:4}) that perturbations as small as \SI{1}{\percent} in alignment of the dipole beam (see the third panel of \cref{fig:states}) rapidly induce these instabilities, resulting in much smother average densities on the left consistent with the experimental images, but performing full high-resolution simulations for direct comparison is prohibitive for this initial study.
We therefore expect that where the simulations produce vortex rings, we can expect to observe solitonic vortices in the experiment.
Despite these radial instabilities, we have verified that using axially symmetric simulations still quantitatively reproduces the bulk dynamics in these elongated systems.

Second, while preparing the \gls{SOC} \gls{BEC} in the experiment, a thermal cloud is generated by the Raman beams in the initial state. 
This is observed in \gls{1D} cross sections of the data.
While in principle one can include the effects of the thermal cloud using the \gls{SPGPE}~\cite{Gardiner:2002, Gardiner:2003, Rooney:2012, Rooney:2014} or \gls{ZNG}~\cite{Zaremba:1999, Allen:2012} formalisms, the agreement between our simulations and experiment show that these effects are small.

\section{SOC Phonon Dispersion}\label{supp:1}
In a previous work, Khamehchi et al.\@ used a single-particle dispersion to describe the expansion of a \gls{SOC} \gls{BEC} into a vacuum, notably matching the speed and \gls{DSW} shape during expansion~\cite{Khamehchi:2017}. 
In this work, the \gls{DSW} is expanding through a non-zero background density. 
From this perspective, \glspl{DSW} are large amplitude phonons described by a phonon dispersion constructed from \gls{BdG} theory.
For a \gls{SOC} \gls{BEC}, this theory predicts linear dispersion for small phonon momenta and a roton-like branch at large phonon momenta. 
This assumes a one-component phonon model. 
To capture the full physics in the regions of negative mass, a two-component model is required.

\begin{figure}[htb]
  \includegraphics[width=\columnwidth]{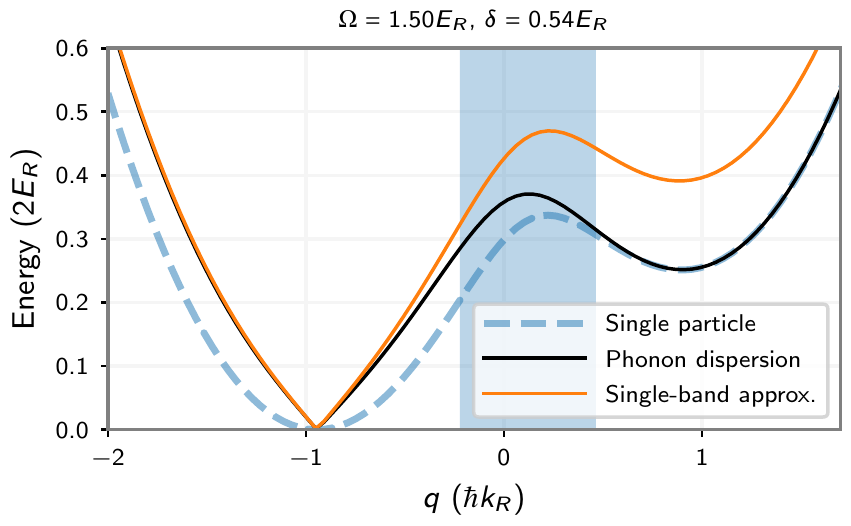}
  \caption{\label{fig:fig6}%
    \textbf{Phonon dispersions with \gls{SOC}.}
    Phonon dispersion (solid) for the full theory (thick black curve) compared with the single-band approximation~\cref{eq:single-band} (thin orange curve) and single-particle dispersion (blue dashed) for experimental parameters $\hbar\Omega = 1.5 E_\mathrm{R}$ and $\hbar\delta = 0.54 E_\mathrm{R}$.
    The blue-shaded area indicates quasimomenta with negative effective mass.
    The phonon dispersion has been shifted to line up with the single particle dispersion minimum for convenience.
    While qualitatively similar, the single-band model has some quantitative differences.
    For this reason, we simulate the full two-component model in all of our results.
}
\end{figure}

In this work, we maintain both components as the single-band model does not correctly reproduce the phonon dispersion in the presence of the superfluid background with \gls{SOC} (see \cref{fig:fig6}). Including the second component shifts the dispersion to a lower energy in the negative mass region, resulting in slower \gls{DSW} propagation in the $+q$-direction.

The two-component, single-band model of the \gls{GPE} is given by
\begin{gather}
  \label{eq:single-band}
  \I\hbar\pdiff{}{t} \ket{\psi} = \left[
    E_{-}(\op{p}) + gn + V_{\text{ext}}(x)
  \right]\ket{\psi},
\end{gather}
where $E_{-}(\op{p})$ is the dispersion of the lower band obtained by diagonalizing equation~\eqref{eq:H2} for homogeneous states.
Here $\ket{\psi}$ is the wavefunction corresponding to the eigenstate of equation~\eqref{eq:H2} describing the lowest band, and is a linear combination of the two bare hyperfine states.
For inhomogeneous densities this picture is locally valid for slowly varying densities, similar to the Thomas-Fermi approximation, and remains valid as long as the system is gently excited compared to the band separation, which is proportional to the strength $\Omega$ of the Raman coupling.
With our parameters, the single-band model exhibits qualitatively similar results to the multi-band description, reproducing many aspects of the experiment, but shows quantitative differences.
The approximate equality of the coupling constants allows one to define a spin-quasimomentum mapping that relates the two-component spin populations $n_\ua$ and $n_\ub$ to the quasimomentum $q$ of the single-component state: 
\begin{gather}
  \label{eq:spin-quasimomentum-map}
  \frac{n_\ub - n_\ua}{n_\ub + n_\ua} = \frac{k - d}{\sqrt{{(k - d)}^2 + w^2}},
\end{gather}
where we have defined the dimensionless parameters $k = p/\hbar k_R$, $d = \delta/4E_R$, and $w = \Omega/4E_R$.
This simplified model captures the interesting phenomena observed in the experiment, but is not quantitatively accurate.

\section{Axially Symmetric GP Simulations}\label{supp:2}

In \cref{fig:comparison} of the main text, experimental results are directly compared to a \gls{1D} Gaussian convolution (mimicking the optical resolution of the experiment) of axially symmetric numerical \gls{GPE} simulations. 
In this figure, the simulations show significant peaks moving to the left that are not observed in the experiments.
This is due to the restricted axial symmetry in the simulations, forcing all defects to be aligned along the imaging axis.
Symmetry unrestricted simulations show that small perturbations cause these defects to align quasi-randomly, averaging out and only the shock fronts remains.
\gls{3D} simulations are computationally expensive to perform for the full system, and while they are able to show an accurate picture of the experimentally observed dynamics, \gls{2D} axially symmetric numerical simulations are able to reproduce some of the key macroscopic features observed in experiments. 

The simulations in \cref{fig:comparison} were made using the axially symmetric GPE with a $N_{x}=8000$ by $N_{r}=50$  grid, a Thomas Fermi radius of $x_{TF}=\SI{150}{\micro\meter}$, and the barrier depths $\SI{-30}{nK}$, $\SI{-60}{nK}$, and $\SI{-90}{nK}$, respectively.

The simulated data in \AxialA, \AxialB, and \AxialC{} show detailed dynamics of the system in three time regimes:
First, from $t_\mathrm{wait}=\SI{-10}{ms}$ to $\SI{0}{ms}$, showing the dynamic generation of turbulent features;
second, from $t_\mathrm{wait}=\SI{0}{ms}$ to $\SI{6}{ms}$, showing the interaction between the \gls{DSW} and the turbulent features stabilizing the right-hand side shock wave;
third, from $t_\mathrm{wait}=\SI{6}{ms}$ onward, showing the motion of the turbulent features once the density peak has passed.

The attractive potential is turned on at $t_\mathrm{wait}=\SI{-10}{ms}$ and the superfluid floods into the barrier, forming a large peak in the center. 
Within $\SI{2}{ms}$, areas of modulated density form grey solitons at the edges of barrier. 
For shallow barriers ($\SI{-30}{nK}$) these solitons are stable.
However, for larger barriers ($\SI{-60}{nK}$ and $\SI{-90}{nK}$) a snaking instability sets in, nucleating vortex rings that appear as vortex anti-vortex pairs in an axial simulations.
The vortex rings form mostly with their central flow oriented away from the center of the cloud.
This is clearly seen in the $\SI{-30}{nK}$ simulation where all the vortex ring have a net outward flow and move in that direction.
For larger potential heights, vortex rings of opposite orientation also form. 

The vortex rings move according to the Magnus relation $\vect{v} \propto \vect{k} \times \vect{F}$, where $\vect{v}$ is the velocity of the vortex, $\vect{k}$ is the circulation of the vortex, and $\vect{F}$ is a force acting on the vortex.
If a vortex ring experiences a force in the same direction as its central flow, it will expand. A vortex ring experiencing a force in the opposite direction to its central flow will shrink.  
In the present setting,
the vortex rings see a flow towards the attractive potential (altering $\vect{v}$) and a density gradient from the density peak ($\vect{F} = -\nabla gn$) which expands rings that are orientated away from density peak and shrinks rings that are orientated towards it. 
For larger potential heights, non-equilibrium dynamics, including vortex ring collisions and annihilations, distort these features.

The attractive potential is turned off in $\SI{0.1}{ms}$ and at $t_\mathrm{wait}=\SI{0}{ms}$ is completely off as the central density peak expands outward, pushing past the turbulent features. 
The density gradient widens (shrinks) vortex rings of same (opposite) central flow. 
If a ring becomes too small, it will annihilate through a Jones-Robert soliton~\cite{Roberts:1971, Jones:1982, Katsimiga:2018, Wang:2019}.
Vortex rings with central flow oriented towards the \gls{SOC} have a larger phase space to annihilate, leading to less vorticity, less effective viscosity, and less dissipation on the right hand side. 

A closer look at the individual components around a vortex ring shows that areas with flow in the direction of the \gls{SOC} have higher densities of $\ket{\ub}$. 
As the density peaks pass through the vortex rings, the bump moving in the direction of the \gls{SOC} converts particles from $\ket{\ua}$ to $\ket{\ub}$, while the bump traveling in the opposite direction  converts particles from $\ket{\ub}$ to $\ket{\ua}$. 
For wait times longer than $t_\mathrm{wait}=\SI{8}{ms}$, the density bumps have developed \gls{DSW} structures and the remaining vortex rings move according to the Magnus relation expected from their flow. 

In \cref{fig:comparison}b at $t_\mathrm{wait}=20$ and $26$~ms, as the large excitation moves through a vortex ring the integrated density appears as a double peaked intensity.
This causes the tracked excitation peak in the top panel of \cref{fig:comparison}b to appear kinked in the numerical (blue shaded) results. 

\section{Axial verses 3D Numerics}\label{supp:3}
Due to the computational memory needed for full \gls{3D} calculations, the \gls{BEC} is modelled in a smaller box, with $L_\mathrm{x}=\SI{120}{\micro\meter}$, $N_{xyz} = (1200, 64, 64)$, and a periodic trapping potential.
This reduces the memory costs by a factor of 4, and is accurate for short times and dynamics in the center of the cloud.
Similarly, in some cases, a quasi-\gls{1D} simulation using techniques like the \gls{NPSE}~\cite{Mateo:2008, Mateo:2009, Mateo:2014, Munoz-Mateo:2014} and \gls{dr-GPE}~\cite{Massignan:2003} can quantitatively reproduce the \gls{3D} dynamics.
However, these are insufficient once features like vortices appear, as shown in~\cite{Lowman:2013a}.

Comparing \gls{3D} to axially-symmetric simulations in \href{https://youtu.be/UcXwBZ7liJE}{Axial 3D tilt \SI{30}{nK}}, we see almost exact agreement during the $\SI{10}{ms}$ long attractive potential pulse. 
This includes the formation of solitons, vorticies pulled in from the boundary, and the snaking creation of vortex rings.
The simulations differ as the density peak splits and pushes past the central vorticies: the \gls{3D} simulations allow for more vorticies to remain.
However, there is agreement between the shape and speed of the \glspl{DSW}, and the speed of solitonic and votex features.
We have verified that \gls{3D} simulations are required to qualitatively explain the observed behavior.

To test the stability of vortex rings against small changes of the experimental parameters, we simulated the attractive dipole beam with a \SI{1}{\percent} tilt in the y-direction, $V_{DB} \propto \exp{((x+0.01y)^2/2\sigma)}$.
In this data (\cref{fig:states} third panel) and supplementary animations, we see that main features like the \gls{DSW} maintain their structure.
However, many of the vortex rings decay in to vortex lines that terminate at cloud edge.

\section{Time-of-Flight dynamics}\label{supp:4}
The technique used to numerically simulate the expansion of the system during the $\SI{10.1}{ms}$ time-of-flight corresponds to setting $\lambda_1(t) = 1$ and $\lambda_2(t) = \lambda_3(t) = \lambda_\perp(t)$ in Eqs.~(11) and (15) of Ref.~\cite{Castin:1996}.
The evolution of dynamics during time-of-flight is shown in \href{https://youtu.be/YO6fEDD-9zI}{Axial \SI{90}{nK} ToF twait \SI{26}{ms}}

When the Raman lasers inducing the \gls{SOC} are switched off, the system is projected into the undressed basis of states $\ket{\ua}$ and $\ket{\ub}$.
Without \gls{SOC} to dress their momenta, the $\ket{\ua}$ component moves slowly to the right while the $\ket{\ub}$ component moves rapidly to the left, making it difficult to locate the center of the cloud.
This can be understood in terms of the background ground state which is a linear combination of mostly $\ket{\ua}$ (with density $n_\ua$) having momentum $k_\ua = k_0 + k_R$ and some $\ket{\ub}$ (with density $n_\ub$) having momentum $k_\ub = k_0 - k_R$.
The spin-quasimomentum map ensures that the background has zero net momentum: $n_\ua k_\ua + n_\ub k_\ub = (n_\ua + n_\ub)k_0 + (n_\ua - n_\ub)k_R = 0$. With our detuning, $n_\ua > n_\ub$ and $k_0 \approx -0.945 k_R$. Thus, during expansion, the two components move in opposite horizontal directions in addition to the vertical separation from the Stern-Gerlach technique.
Within the first $\SI{2}{ms}$ of expansion the cloud expands rapidly, dropping the density by a factor of \num{10}. 
The most notable features after expansion come from low density objects such as solitons, vortex rings, and \gls{DSW} where areas with these features deepen and widen, pushing density to either side.
The density pile-up of nearby vortex rings will often constructively interfere, resulting in some of the largest peaks during expansion dynamics.

\bibliography{ThesisRef,master,local}

\end{document}